\documentclass[preprint,noshowkeys,amsmath,amssymb,superscriptaddress,nofootinbib]{revtex4}

\usepackage{bm}
\usepackage{slashed}

\usepackage[pdftex]{graphicx,color}
\usepackage{epstopdf}
\usepackage[bookmarks=true,bookmarksnumbered=true,bookmarkstype=toc, colorlinks=true, citecolor=blue, linkcolor=blue]{hyperref} 
\usepackage{units}

\usepackage{mathtools}

\usepackage{color}

\definecolor{green}{cmyk}{1,0,1,0}

\definecolor{pink}{cmyk}{0,0.5,0,0}

\definecolor{pastelpink}{cmyk}{0,0.25,0,0}

\definecolor{softpink}{cmyk}{0,0.125,0,0}

\definecolor{purple}{cmyk}{0.5,1.0,0.1,0}

\definecolor{violet}{cmyk}{0.75,1,0.25,0}

\usepackage[utf8]{inputenc}

\preprint{UME-PP-013}
\preprint{EPHOU-20-009}

\begin{document}

\title{Signal from sterile neutrino dark matter in extra $U(1)$ model \\ at direct detection experiment}

\author{Osamu Seto}
\email{seto@particle.sci.hokudai.ac.jp}
\affiliation{Institute for the Advancement of Higher Education, Hokkaido University, Sapporo 060-0817, Japan}
\affiliation{Department of Physics, Hokkaido University, Sapporo 060-0810, Japan}

\author{Takashi Shimomura}
\email{shimomura@cc.miyazaki-u.ac.jp}
\affiliation{Faculty of Education, Miyazaki University, Miyazaki, 889-2192, Japan}

\begin{abstract}
We examine the possibility that direct dark matter detection experiments find decay products from sterile neutrino dark mater 
in $U(1)_{B-L}$ and $U(1)_R$ models. 
This is possible if the sterile neutrino interacts with a light gauge boson, and decays into a neutrino and the light gauge boson with a certain lifetime.
This decay produces energetic neutrinos scattering off nuclei with a large enough recoil energy in direct dark matter detection experiments.
We stress that direct dark matter detection experiments can explore not only WIMP but also sterile neutrino dark matter.
\end{abstract}

\date{\today}

\maketitle

\section{Introduction}

Direct detection of dark matter (DM) aims to prove the existence of DM by discovery and reveal identity by examining the property through their scattering off a target in underground detectors. Over the past years, the LHC and many direct detection experiments have reported null results for a substantial portion of the expected parameter space for Weakly Interacting Massive Particle (WIMP) DM \cite{Agnese:2017njq,Aprile:2018dbl}. Thus, an interesting alternative to the WIMP hypothesis have been received a lot of attention recently.

Not only the existence of non-baryonic DM but also non-vanishing neutrino masses are regarded as the most clear evidences for new physics beyond the standard model (SM). 
One of the simplest explanations of tiny neutrino masses is the so called seesaw mechanism 
with gauge singlet right-handed (RH) neutrinos~\cite{Minkowski:1977sc,Yanagida:1979as,GellMann:1980vs,Mohapatra:1979ia}.
Two neutrino mass differences measured in solar and atmospheric neutrino oscillations can 
be explained with two RH neutrinos.
Nevertheless, as other SM fermions, three generations of  RH neutrinos has been often assumed, 
because this assumption looks natural.
Then, one RH neutrino could be DM candidate if its lifetime is long enough, because it is electrically neutral~\cite{Boyarsky:2018tvu}.
Thus, three RH neutrino extension of the SM is simple and economical from 
the viewpoint of the simultaneous explanation of tiny neutrino mass and DM.
This kind of model is called the $\nu$MSM~\cite{Asaka:2005an,Asaka:2005pn}.

Three generations can be theoretically verified in a simple extension of the SM model, 
once the SM is extended by introducing an extra $U(1)$ gauge symmetry and RH neutrinos 
are charged under this symmetry. Typical examples are $U(1)_{B-L}$~\cite{Davidson:1978pm,Mohapatra:1980qe,Marshak:1979fm} and $U(1)_R$~\cite{Jung:2009jz}.
In such a model, the presence of the three RH neutrinos is theoretically required 
for the cancellation of the gauge and mixed gauge-gravitational anomalies \cite{Ko:2013zsa,Nomura:2017tih}.

An extra $U(1)$ interaction is also beneficial to consistent realization of  sterile neutrino DM.
On the one hand, for gauge singlet RH neutrinos, sterile neutrinos as dark matter are 
generated in the early Universe by so-called Dodelson-Widrow (DW) mechanism through 
the mixing between active and sterile neutrinos~\cite{Dodelson:1993je,Dolgov:2000ew}. 
However, the stringent bound from X-ray background excludes a large enough mixing for 
the DW mechanism~\cite{Asaka:2006rw,Yuksel:2007xh}.
On the other hand, if RH neutrinos interact through an extra $U(1)$ gauge interaction, 
sterile neutrino DM can be generated nonthermally.
One kind of nonthermal productions, called ``freeze-in mechanism'', of sterile neutrino DM was first considered 
for the heavy extra gauge boson under cosmology with a low reheating temperature in Ref.~\cite{Khalil:2008kp}
and later considered for the light extra gauge boson in Ref.~\cite{Kaneta:2016vkq,Biswas:2016bfo}.

Sterile neutrino DM has been probed by astrophysical observations such as X-ray background 
as mentioned above.
The X-ray signal from the sterile neutrino decay into a lighter neutrino and a photon is 
an unique process to explore sterile neutrino DM.
An extra $U(1)$ interacting sterile neutrino may decay into a lighter neutrino and the extra gauge boson.
This decay mode opens a new channel to probe sterile neutrino DM.
We point out that detection of light neutrinos as the decay product is possible 
by ongoing and future experiments of direct dark matter detection \cite{Aalbers:2016jon,Aalseth:2017fik,Akerib:2018lyp}. 
While such experiments are primarily aimed to detect WIMP DM,
we show that those are also capable to detect a signal from sterile neutrino decay.

\section{Model} 
\label{sec:model}

\begin{table}[t]
  \begin{center}
    \begin{tabular}{|c||cccccc||cc|} \hline
	& ~~~$Q$~~~ & ~~~$u_R$~~~ & ~~~$d_R$~~~ & ~~~$L$~~~ & ~~~$e_R$~~~ & 
	~~~$\nu_R$~~~ & ~~~$H$~~~ & ~~~$S$~~~  \\ \hline \hline
	$SU(3)$ & $3$ & $3$ & $3$ & $1$ & $1$ & $1$ & $1$ & $1$ \\ \hline
	$SU(2)_L$ & $2$ & $1$ & $1$ & $2$ & $1$ & $1$ & $2$ & $1$ \\ \hline
	$U(1)_{Y}$ & $\frac{1}{6}$ & $\frac{2}{3}$ & $-\frac{1}{3}$ & $-\frac{1}{2}$ & $-1$ & $0$ 
	     & $\frac{1}{2}$ & $0$\\ \hline \hline
	$U(1)_{B-L}$ & $\frac{1}{3}$ & $\frac{1}{3}$ & $\frac{1}{3}$ & $-1$ & $-1$ & $-1$ & $0$ & $+2$ \\ \hline
	$U(1)_R$ & $0$ & $\frac{1}{2}$ & $-\frac{1}{2}$ & $0$ & $-\frac{1}{2}$ & $\frac{1}{2}$ & $\frac{1}{2}$ & $-1$ \\ \hline
    \end{tabular}
  \end{center}
\caption{Matter contents and charge assignment of the fields. }
\label{tab:matter-contents}
\end{table}

We consider the extension of the SM by imposing $U(1)_{B-L}$ or $U(1)_R$ gauge symmetry, where $B$ and $L$ are the baryon and 
lepton number, and $R$ represents right-handed chirality, respectively. 
As we mentioned in the introduction,  the fermion sector of the SM is extended with three 
RH neutrinos $\nu_R$. The scalar sector is also extended by introducing at least, one complex scalar $S$, which is charged 
under the extra $U(1)$ symmetry. The charge assignment of the particles are given in Table~\ref{tab:matter-contents}.

The kinetic terms of the model with the $U(1)_X$ symmetry ($X=B-L$ or $R$) are given by
\begin{align}
\mathcal{L}_{\mathrm{kin.}} = i \overline{f} \slashed{D} f + |D_\mu H|^2 + |D_\mu S|^2 
- \frac{1}{4} \tilde{W}_{\mu\nu} \tilde{W}^{\mu \nu}-\frac{1}{4} \tilde{B}_{\mu\nu} \tilde{B}^{\mu \nu} 
- \frac{1}{4} \tilde{X}_{\mu\nu} \tilde{X}^{\mu \nu} , \label{eq:lag-kin}
\end{align}
with the covariant derivative $D_\mu = \partial_\mu - i g_2 \tilde{W}_\mu - i Y_f g_1 \tilde{B} - i x_f g_X \tilde{X}_\mu$.
Here, $\tilde{W},~\tilde{B}$ and $\tilde{X}$ represent the gauge fields in the interaction basis and 
$g_2,~g_1$ and $g_X$ are the gauge coupling constants of $SU(2)_L,~U(1)_Y$ and $U(1)_X$, respectively.  
The fermion $f$ denotes $Q,~u_R,~d_R$ and $L,~e_R,~\nu_R$ with
$Y_f$ and $x_f$ being the $U(1)_Y$ and $U(1)_X$ charges, respectively. 
We omit any symbol about the $SU(3)$ color interaction. 
The gauge kinetic mixing term for two  $U(1)$ gauge fields is allowed by the symmetry, however 
we drop this term in our analysis for simplicity. The results are unchanged as long as the mixing parameter 
is smaller than $g_X$.\footnote{Even though gauge kinetic mixing parameter is vanishing at tree-level, 
it can be generated via loop effects. At one-loop level, the loop-induced kinetic mixing paramter is given by 
$\epsilon_{\mathrm{loop}} \sim e g_X/(4 \pi^2) \sum Q_{\mathrm{em}} Q_X$, where $Q_{\mathrm{em}}$ 
and $Q_X$ are the 
electric charge and $U(1)_X$ charge, respectively. This is $10^{-3}$ times smaller than $g_X$, 
and hence the loop-induced mixing can also be ignored compared to $g_X$.}
The Yukawa interactions are given as
\begin{align}
\mathcal{L}_{\mathrm{yukawa}} &= Y_u \overline{Q} \tilde{H} u_R + Y_d \overline{Q} H d_R  
+ Y_e \overline{L} H e_R \nonumber \\
&\quad + Y_\nu \overline{L} \tilde{H} \nu_R  + Y_R \overline{\nu_R^c} S \nu_R + h.c. , \label{eq:lag-yukawa}
\end{align}
where the Dirac Yukawa matrices are denoted as $Y_u,~Y_d$ and $Y_e,~Y_\nu$ 
for up, down quarks and charged leptons, neutrinos, respectively. 
The Yukawa matrix for $\nu_R$ is denoted as $Y_R$. 
Here $\tilde{H}$ represents $i\sigma_2 H^\ast$ where $\sigma_2$ is the Pauli matrix. 
Note that flavour and generation indices are omitted for simplicity.

After the scalar fields develop the vacuum expectation value (vev) $\langle H \rangle^T = (0, v)/\sqrt{2}$ and 
$\langle S \rangle = v_s/\sqrt{2}$, with $v \simeq 246$ GeV, the fermions and gauge bosons acquire the masses. 
Neutrinos obtain the Majorana and Dirac mass terms, and then the masses and mixing angles of three active neutrinos 
can be reproduced by the type-I seesaw mechanism with two RH neutrinos. 
We identify the remaining RH neutrino, suppose $\nu_{R_1}$, as the sterile neutrinos DM $\nu_s$ with 
tiny active-sterile mixing, $\theta$, 
\begin{align}
\nu_s = \nu_{R_1} + \sin\theta U_{1\alpha} \nu_{L\alpha}^c,
\end{align}
where $U$ is the MNS matrix. The superscript $c$ stands for charge conjugation, and the index $\alpha$ runs over $e,~\mu,~\tau$.

Then, the interaction Lagrangian of the $X$ boson and fermions in mass eigenstates takes the form of 
\begin{align}
\mathcal{L}_{\mathrm{int.}} = i e\overline{f} \gamma^\mu ( \epsilon_f^V + \epsilon_f^A \gamma_5) f X_\mu,
\label{eq:lag-x-f-int}
\end{align}
where $\epsilon_f^V$ and $\epsilon_f^A$ are the vector and axial vector coupling defined by
\begin{align}
\epsilon_f^V &= \frac{1}{2} (x_{f_R} + x_{f_L}) \epsilon_X \cos\chi  - \left( \frac{1}{2} T_f - Q_f \sin^2\theta_W \right) \epsilon_{\mathrm{NC}}, \label{eq:eps-f-v}\\
\epsilon_f^A &= \frac{1}{2} (x_{f_R} - x_{f_L}) \epsilon_X \cos\chi  + \frac{1}{2} T_f \epsilon_{\mathrm{NC}}.
\label{eq:eps-f-a}
\end{align}
Here $e$ is the elementary charge and $\epsilon_X$ is defined as $g_X/e$. The weak isospin and electric charge 
are denoted as $T_f$ and $Q_f$, respectively, and $\theta_W$ is the Weinberg angle.
The mixing angle $\chi$ relates the interaction-eigenstates $(\tilde{Z},~\tilde{X})$ with the mass-eigenstates  $(Z,~X)$ as 
\begin{align}
\begin{pmatrix}
\tilde{Z} \\
\tilde{X}
\end{pmatrix}
=
\begin{pmatrix}
\cos\chi & -\sin\chi \\
\sin\chi & \cos\chi
\end{pmatrix}
\begin{pmatrix}
Z \\
X
\end{pmatrix},
\end{align}
where the mixing angle is defined by
\begin{align}
\tan\chi = - 2 \epsilon_X Q_{X,H} \cos\theta_W \sin\theta_W \frac{m_Z^2}{m_Z^2 - m_X^2 }.
\end{align}
Here, $Q_{X,H}$ is the $U(1)_X$ charge of $H$ and $\tilde{Z}$ is the $Z$ boson in the SM. 
The neutral current contribution $\epsilon_{\mathrm{NC}}$ is given\footnote{
For the $U(1)_{B-L}$ scenario, the loop-induced gauge mixing  
gives $\epsilon_{NC} \simeq 10^{-3} \epsilon_X$. In Eqs.~\eqref{eq:eps-f-v} and \eqref{eq:eps-f-a}, 
such $\epsilon_{NC}$ is negligible compared with the first terms. Thus, we can safely ignore the effects of the loop-induced kinetic mixing.
}
\begin{align}
\epsilon_{\mathrm{NC}} = \frac{\sin\chi}{\sin\theta_W \cos\theta_W}.
\end{align}
Here, we note one important difference between $U(1)_{R}$ model and $U(1)_{B-L}$ model with the minimal Higgs sector.
That is, at tree-level, the $Z$ boson and the $X$ boson do not mix without the presence of gauge kinetic mixing in the minimal $U(1)_{B-L}$ model,
while in the minimal $U(1)_R$ model the $Z$ boson and the $X$ boson mix even for vanishing gauge kinetic mixing, because the $SU(2)$ doublet Higgs field has to be charged under the $U(1)_R$ symmetry and generates the mass mixing between the $Z$ boson and the $X$ boson~\cite{Seto:2020jal}.  
In addition to Eq.~\eqref{eq:lag-x-f-int}, the sterile neutrino DM has the interaction with $X$ and the active neutrinos 
$\nu_i~(i=1,2,3)$ given by
\begin{align}
\mathcal{L} = i e \sin\theta U_{1\alpha} \delta_{\alpha i} \overline{\nu^c_s} 
\gamma^\mu ( \epsilon^V - \epsilon^A \gamma_5) \nu_i X_\mu 
+ h.c.,
\end{align}
where $\epsilon^V = -\epsilon_{\nu_L}^V + \epsilon_{\nu_R}^V,~\epsilon^A = \epsilon_{\nu_L}^A + \epsilon_{\nu_R}^A$, respectively.

\section{Sterile neutrino dark matter} 
\label{sec:darkmatter}

\subsection{Cosmological abundance}

Throughout this paper, we consider only the case of $2 m_{\nu_s} > m_X$, where $m_X$ is the mass of the $X$ boson. 
For this mass spectrum, the resultant DM abundance generated by freeze-in mechanism has interesting feature. That is independent from the sterile neutrino DM mass and almost determined by the extra $U(1)$ gauge coupling only~\cite{Kaneta:2016vkq}.

Freeze-in production of sterile neutrino $\nu_s$ with the mass $m_{\nu_s}$ is governed by the Boltzmann equation 
\begin{align}
 \frac{d n_{\nu_s}}{dt}+ 3 H n_{\nu_s} & = C_\mathrm{coll.}(f \bar{f} \rightarrow \nu_s \nu_s)  \nonumber \\
 & = \sum_f \langle \sigma v(f \bar{f} \rightarrow \nu_s \nu_s)\rangle n_f n_{\bar{f}} ,
\label{eq:boltzmann}
\end{align}
where $n_{\nu_s}$ is the number density of DM sterile neutrino $\nu_s$, $H$ is the Hubble parameter, 
$\langle \sigma v\rangle$ is the thermal averaged product of the cross section and the relative velocity 
for production processes and $n_{f(\overline{f})}$ is the thermal abundance of a SM fermion $f(\overline{f})$ in initial states.
The present DM abundance is estimated by integrating Eq.~(\ref{eq:boltzmann}) as 
\begin{align}
 \Omega_{\nu_s} h^2 & = 2.82\times 10^3 \frac{m_{\nu_s}Y_{\nu_s}}{\mathrm{eV}} ,\\
 Y_{\nu_s} & = \int^{T_R}_{T_0} \frac{C_\mathrm{coll.}}{s H T} dT,
\end{align}
where $s$ is the entropy density, $T_R$ is the reheating temperature after inflation and $T_0$ is a low temperature.
The remarkable feature of this is, as is well known, that the final abundance is independent from $T_R$ and the mass $m_{\nu_s}$. 
By substituting 
\begin{align}
H M_P = \sqrt{\frac{\pi^2 g_*}{90}} T^2, s =\frac{2\pi^2}{45}g_* T^3, \langle \sigma v\rangle \sim \frac{g_X^4}{T^2},
\end{align}
with $g_*$ being the relativistic degrees of freedoms, $M_P$ being the reduced Planck mass, 
and $g_X$ is the gauge coupling constant, one easily finds that the production is efficient only at $T\simeq m_{\nu_s}$ and obtain $Y \propto 1/m_{\nu_s}$.
Kaneta et al performed the detailed calculation of this freeze-in production for the same model 
and obtained ~\cite{Kaneta:2016vkq}
\begin{align}
\Omega_{\nu_s} h^2 \simeq 0.12\left(\frac{g_X}{4.5\times 10^{-6}}\right)^4 .
\end{align}
If $g_X$ is smaller, then $\nu_s$ becomes a sub-dominant component of the total DM.
In the rest of the paper, we fix the value of coupling as above
to reproduce the appropriate DM abundance, unless stated.

\subsection{Constraints}

\subsubsection{The $X$ boson mass}

One stringent constraint on a light gauge boson comes from the $e-\nu$ scattering experiments 
such as GEMMA~\cite{Beda:2013mta}, BOREXINO~\cite{Agostini:2018uly} and TEXONO~\cite{Deniz:2010mp}.  
Various beam dump experiments such as CHARM~\cite{Bergsma:1985is,Gninenko:2012eq} and NOMAD~\cite{Astier:2001ck}
and electron colliders such as KLEO~\cite{Anastasi:2015qla} and BABAR~\cite{Lees:2014xha} constrain wide range of 
the $X$ boson mass for $m_X > 2 m_e$ where $X$ decays into an electron and positron pair~\cite{Lindner:2018kjo}.
The $X$ boson decaying into electron and positron are constrained by beam dump experiments.
For $g_X = \mathcal{O}(10^{-6})$, currently $1 \, \mathrm{MeV} < m_X \lesssim 100$ MeV are excluded.
$m_X < 1$ MeV is not excluded by beam damp experiments, however, $m_X \lesssim 0.3$ MeV is excluded 
by the stellar cooling in globular clusters~\cite{Grifols:1986fc,Harnik:2012ni,Bilmis:2015lja}.
The BBN constraint is weaker than the bound by BOREXINO 
for the mass spectrum $2 m_{\nu_s} > m_X$. See for instance Ref.~\cite{Kaneta:2016vkq}.
The unconstrained mass range of the $X$ boson are 
\begin{align}
& 0.3 \,\mathrm{MeV} \lesssim m_X < 1 \,\mathrm{MeV},
\label{ineq:band} \\ 
& 100 \,\mathrm{MeV} \lesssim m_X  
\label{ineq:heavy},
\end{align}
for $g_X = \mathcal{O}(10^{-6})$.
For a smaller $g_X$ where sterile neutrinos become sub-dominant components of DM, 
the lower bound from the beam dump experiments increases.

\subsubsection{Lifetime and cosmic ray background bound}

For sterile neutrino to be DM, its life time must be longer than the age of our Universe.
If kinematically allowed, 
the sterile neutrino decays into one neutrino and the $X$ boson through active-sterile mixing. 
The decay rate for this main mode of the sterile neutrino is given by 
\begin{align}
\Gamma(\nu_s \rightarrow X \nu) = \frac{ e^2 (\epsilon^V )^2 \sin^2\theta}{\pi} m_{\nu_s}
\left( 1 - \frac{m_X^2}{m_{\nu_s}^2} \right) \left( \frac{m_{\nu_s}^2}{m_X^2} + 1 - 2 \frac{m_X^2}{m_{\nu_s}^2} \right),
\end{align}
which is much larger than the SM contribution~\cite{Pal:1981rm}.
For a very tiny mixing $\theta$, the lifetime can be long enough.

In fact, the constraints from X-ray background is more stringent for sterile neutrino DM than that from the lifetime. 
The decay rate from the SM processes is given as~\cite{Pal:1981rm}
\begin{align}
\Gamma(\nu_s \rightarrow \gamma \nu) = \frac{9 \alpha_{\mathrm{em}} G_F^2}{256 \pi^4} m_{\nu_s}^5 \sin^2\theta ,
\end{align}
while that from the $X$ boson mediation vanishes due to the electromagnetic gauge symmetry.
If the $X$ boson is heavier than $2m_e$, $\nu_s$ decays into $\nu$ and $X$, followed by 
the decay of $X$ boson into $e^+ e^-$, as $\nu_s \rightarrow \nu X \rightarrow \nu e^- e^+$.
Although interstellar sub-GeV electrons and positrons can be hardly detected because of nearby magnetic field, Boudaud et al~\cite{Boudaud:2016mos} pointed out that Voyager I data~\cite{Cummings:2016pdr} constrains annihilation or decay of a DM particle which injects energetic electrons and positrons $E_{e^{\pm}} > 8$ MeV.
Thus, unless the lifetime, $\tau_{\nu_s}=1/\Gamma(\nu_s \rightarrow X \nu)$, is longer than $10^{27}$ second, 
we have the upper limit as $m_X \lesssim 16$ MeV 
which is obviously conflict with beam dump experiment bound (\ref{ineq:heavy}).
After all, we will consider the mass band (\ref{ineq:band}) only for our purpose and will take $m_X=0.5$ MeV.
For information, here we note $v_s \sim m_X/g_X \sim 10^2$ GeV.

\begin{figure}[t]
\begin{tabular}{ccc}
& \includegraphics[width=0.8\textwidth]{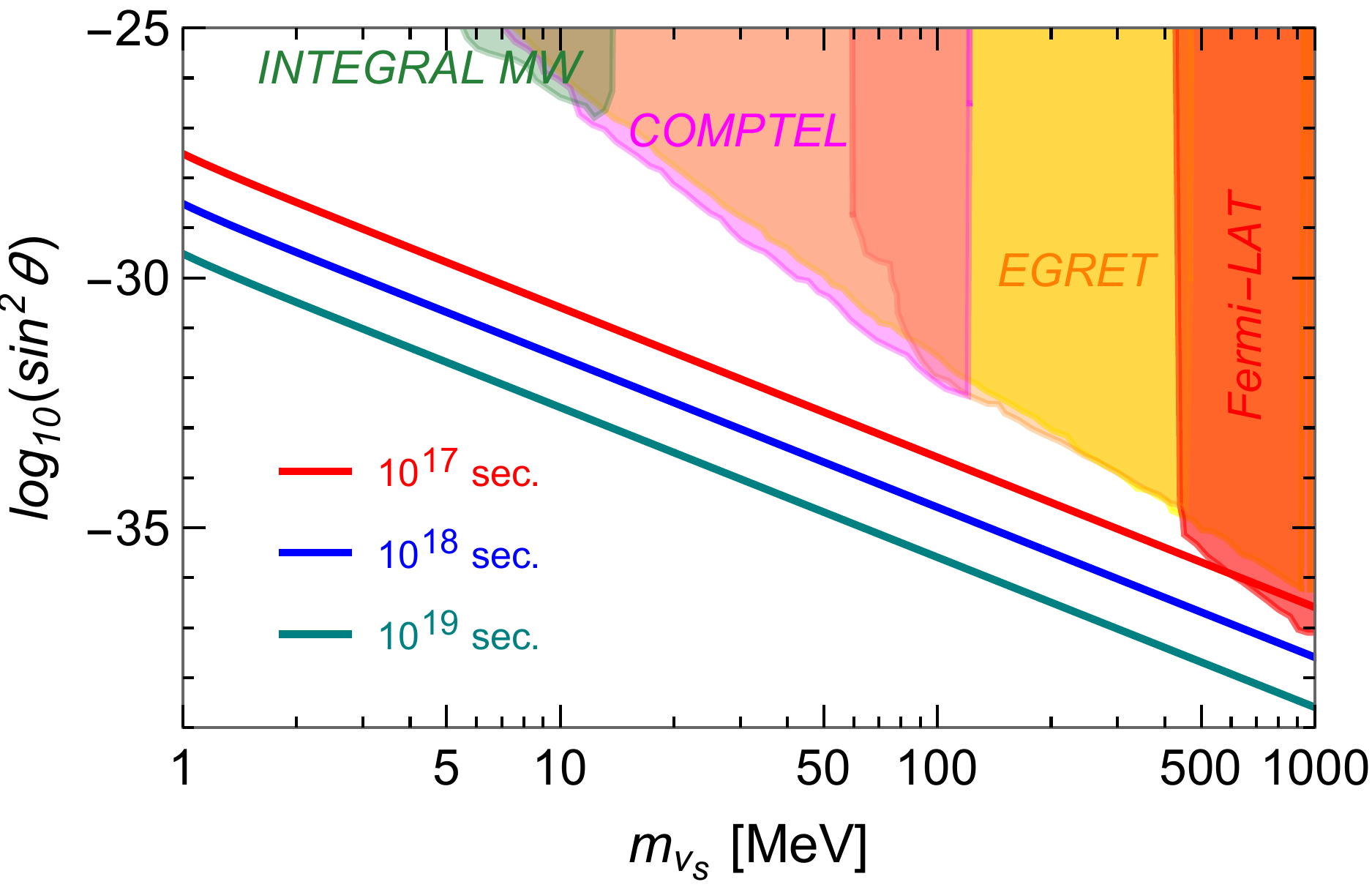} &
\end{tabular}
\caption{The sterile DM lifetime and astrophysical constraints on the active-sterile mixing and the sterile DM mass. }
\label{fig:mixing-mass}
\end{figure}
Figure \ref{fig:mixing-mass} shows the left-right mixing as a function of $m_{\nu_s}$. The parameters are 
taken as $\epsilon^V = 4.5 \times 10^{-6}/e$, $m_X = 0.5$ MeV. The red, blue and green lines correspond to 
the lifetime of the sterile neutrino DM, $10^{17},~10^{18}$ and $10^{19}$ sec., respectively. Color filled regions are excluded by 
Fermi-Lat (red), EGRET (orange), COMPTEL (magenta), and INTEGRAL MW (green) taken from \cite{DeRomeri:2020wng}.
In the following analysis in this section, we use the lifetime 
$\tau$ as an input parameter instead of the mixing $\theta$.

\subsection{Detection of neutrinos from sterile neutrino decay}

The flux of light active neutrino has two sources. One comes from by sterile neutrino decay $d \phi^{\nu_s}/d E_{\nu}$ and the others, we call the background, include solar, atmospheric and diffuse supernova neutrinos $d \phi^{\mathrm{bkg}}/d E_{\nu}$. The resultant flux is given by
\begin{align}
\frac{d \phi}{d E_{\nu}} &= \frac{d \phi^{\mathrm{bkg}}}{d E_{\nu}}+\frac{d \phi^{\nu_s}}{d E_{\nu}}, \\
\frac{d \phi^{\nu_s}}{d E_{\nu}} &= \frac{d N}{d E_{\nu}}\frac{J}{4 \pi \tau_{\nu_s} m_{\nu_s}} , \label{eq:phiDMnu} \\
J &= \int_\mathrm{l.o.s} ds \rho_{\mathrm{DM}} d\Omega, 
\end{align}
where $J$ is the so-called $J$ factor for decaying DM with the integration along the line of sight and over the solid angle.
We consider the whole Milky Way galaxy halo and, in this paper, 
quote the value $J= 7\times 10^{22}$ GeV/cm${}^2$ from Ref.~\cite{Buch:2020mrg}.
For the decays of the sterile neutrino DM and the $X$ boson mentioned above, the energy spectrum in generated neutrino $dN/dE_{\nu}$ depends on the mass spectrum.
If the mass of $\nu_s$ and $X$ are degenerate, one is given by $E_{\nu} = m_{\nu_s}-m_X$ and the others is given by $E_{\nu} \simeq m_X/2$.
On the other hand, if the $X$ boson is much lighter than $\nu_s$, one is given by $E_{\nu} \simeq m_{\nu_s}/2$ and 
the others from from the X boson decay takes "the box-shape" spectrum~\cite{Ibarra:2012dw,Agashe:2012bn,Boddy:2016fds} as $dN/dE_{\nu} \simeq 2/m_{\nu_s}$ for
$m_X^2/(2m_{\nu_s}) < E_{\nu} < m_{\nu_s}/2$.
If $\tau_{\nu_s}$ is shorter than the age of Universe and the sterile neutrino is a sub-dominant component with the fraction $f_{\nu_s}=\rho_{\nu_s}/\rho_{\mathrm{DM}}$, then $1/\tau_{\nu_s}$ in Eq.~(\ref{eq:phiDMnu}) should be reinterpreted as $f_{\nu_s}/\tau_{\nu_s}$.

Those neutrinos with monochromatic spectrum can be detected in neutrino detection experiments, 
principally, Super-Kamiokande (SK),
the sharp line-like feature is smeared out due to the energy resolution of detectors nevertheless. 
Searches for diffuse supernova neutrino (DSN) through inverse beta decay give the upper bound
on extra flux of $\bar{\nu}_e$~\cite{Bays:2011si,Zhang:2013tua}.
Measurements of atmospheric neutrino flux also provide the upper bound on an extra contribution from 
other neutrino sources~\cite{Richard:2015aua}. 

\begin{figure}[t]
\begin{tabular}{ccc}
& \includegraphics[width=0.8\textwidth]{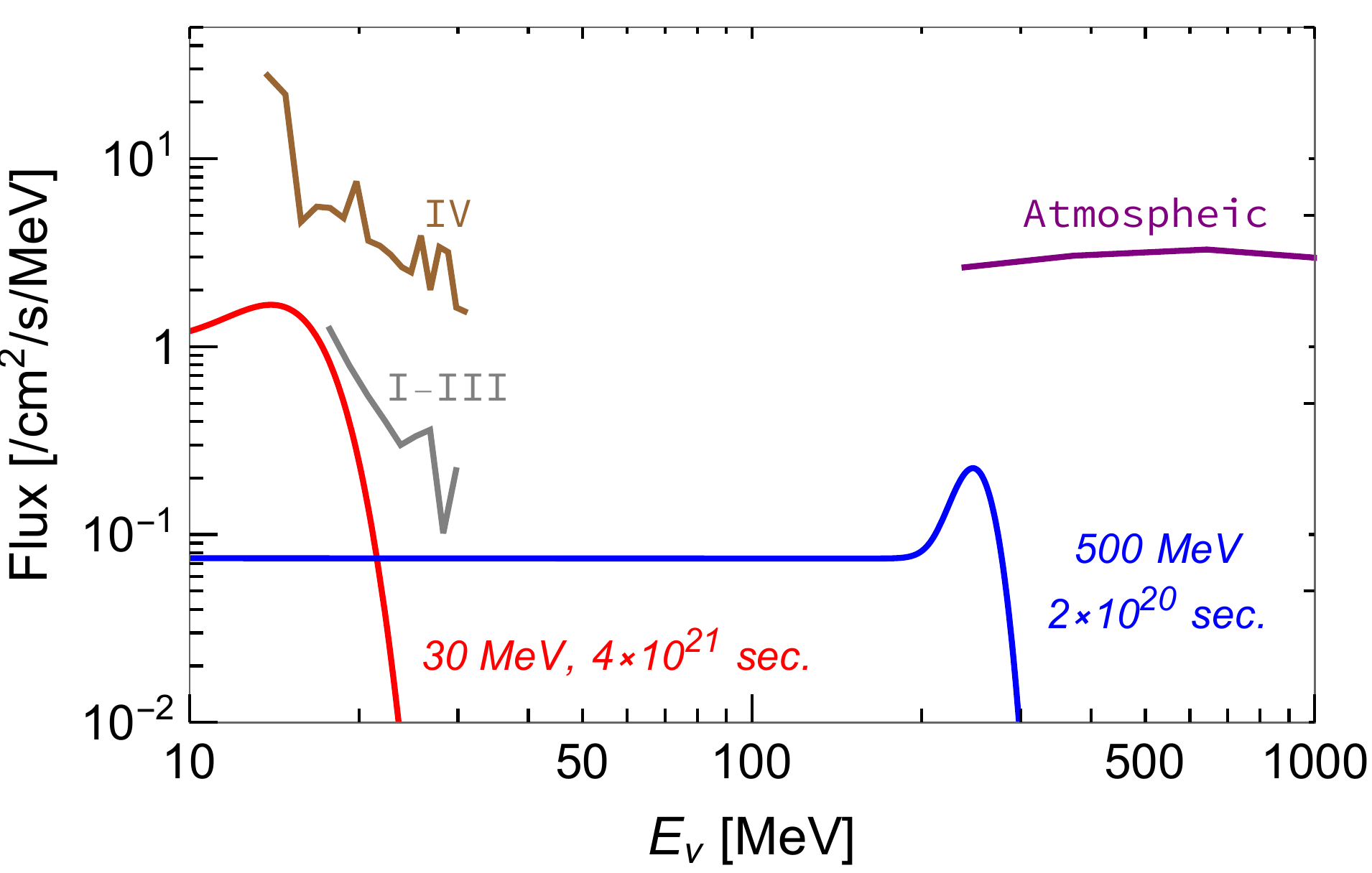} &
\end{tabular}
\caption{Comparison between expected flux for some benchmark points of sterile neutrino DM decay and bounds from null results of DSN searches~\cite{Bays:2011si,Zhang:2013tua} as well as the measurement of atmospheric neutrino~\cite{Richard:2015aua} at SK.}
\label{fig:sk-bound}
\end{figure}
Figure \ref{fig:sk-bound} shows the limits on the neutrino flux by SK and predicted neutrino flux for 
some benchmark points of sterile neutrino DM.
The limits come from the null results in DSN from SK I/II/III data (gray)~\cite{Bays:2011si}, 
SK IV (brown)~\cite{Zhang:2013tua} and the measurement of atmospheric neutrino 
(purple)~\cite{Richard:2015aua} . 
The red and blue curves are predicted neutrino spectrum for some benchmark point and
the values associated with each curves stand for the mass and the lifetime of the sterile neutrino DM.
The spectrum with one sharp peak originated from $E_\nu \simeq m_{\nu_s}/2$ and continuous component from 
the $X$ decay is a prediction of the model.
The plotted flux is evaluated by Eq.~(\ref{eq:phiDMnu}) with replacing $\delta(E_\nu - m_{\nu_s}/2)$ 
with the Gaussian function used in SK~\cite{Cravens:2008aa} to take the energy resolution into account, and multiplying one-sixth because only $\bar{\nu}_e$ is responsible for DSN detection in SK.
For $\mathcal{O}(10)$ MeV sterile neutrino DM mass, its lifetime must be longer than about 
$10^{22}$ second due to stringent limits from DSN search in this energy range. 
On the other hand, shorter lifetime is allowed for $m_{\nu_s} > \mathcal{O}(100)$ MeV.

Next, we consider detection in the direct dark matter detection experiments.
The event rate of recoils is expressed as
\begin{equation}
 \frac{d R}{d E_r} = N_T 
 \int_{E_{\nu}^{\mathrm{min}}}^{E_{\nu}^{\mathrm{max}}} \frac{d \phi}{d E_{\nu}}\frac{d \sigma}{d E_r}dE_{\nu} ,
\end{equation}
where $N_T$ is the number of the total target particles such as nucleus or electrons, 
$d\sigma/dE_r$ is the differential cross section with respect to the recoil energy, 
$E_{\nu}^{\mathrm{max}}$ is the maximal energy in the neutrino flux, and 
$E_{\nu}^{\mathrm{min}}$ is the minimal energy of neutrino to generate a given recoil energy $E_r$ which is  
\begin{equation}
E_{\nu}^{\mathrm{min}} = \sqrt{\frac{m_N E_r}{2}},
\end{equation}
for a scattering with a nucleus with the mass $m_N$.

The differential cross section for the scattering of neutrino and nuclei $N$ with the mass $m_N$ is given by~\cite{Freedman:1973yd}
\begin{align}
\left(\frac{d \sigma}{dE_r}\right)(\nu N \rightarrow \nu N) = \frac{G_F^2}{8 \pi E_{\nu}^2} ((A-Z)-(1-4s_W^2)Z)^2 m_N ( 2E_{\nu}^2 -m_N E_r) F^2(E_r), \label{eq:crosssection:nu-N}
\end{align}
with $F^2(E_r)$ being the nuclear form factor. 
Here, $Z$ and $A$ are the atomic number and the mass number of nuclei, respectively.
The $X$ boson exchange process becomes significant in $d \sigma/dE_r$ only if the gauge coupling is as large as $\mathcal{O}(10^{-4})$ ~\cite{Boehm:2018sux}. For our interest, the $X$ boson exchange processes are negligible. 

\begin{figure}[hbt]
\begin{tabular}{ccc}
& \includegraphics[width=0.8\textwidth]{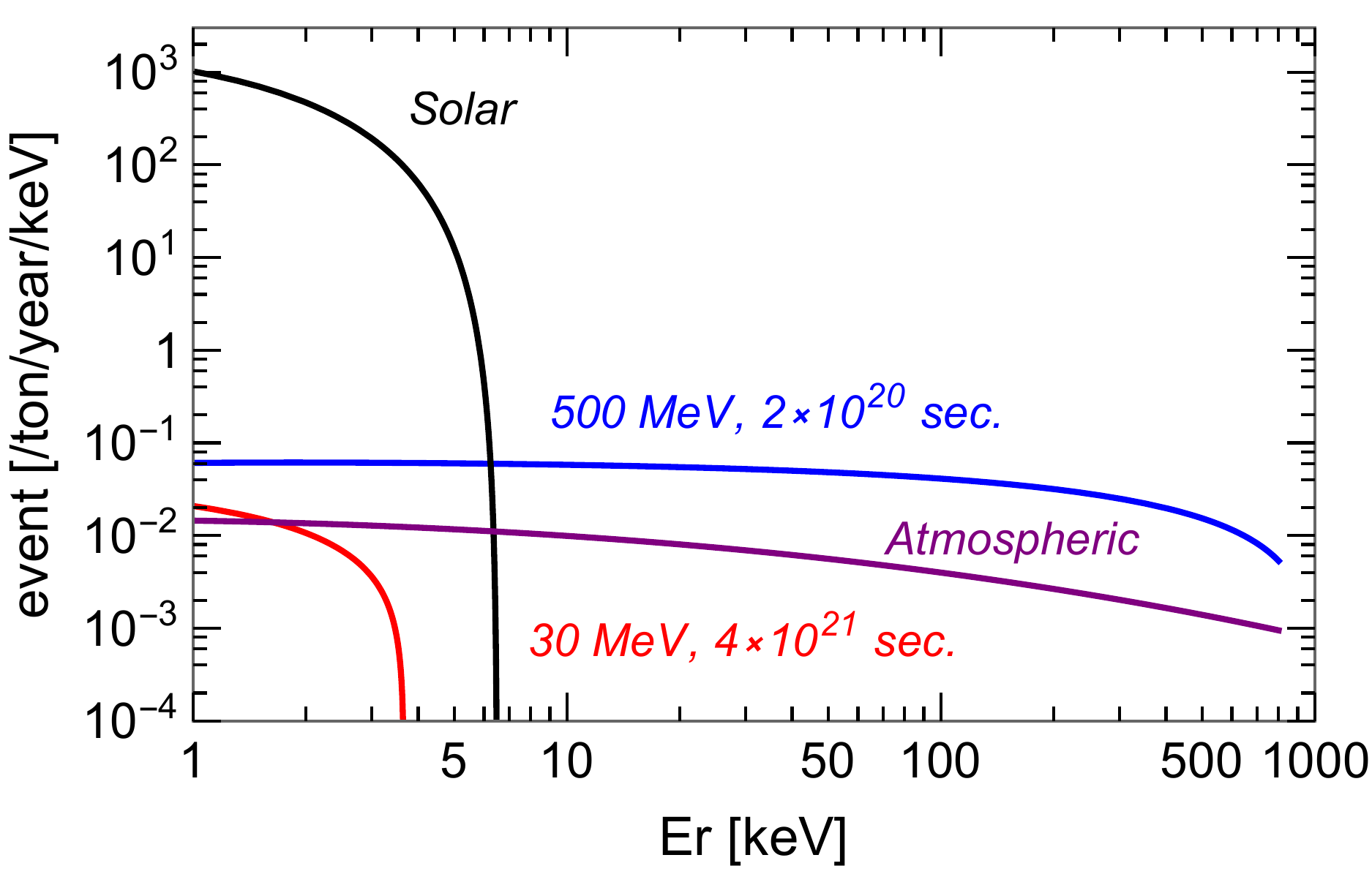} &
\end{tabular}
\caption{The nucleus recoil energy spectrum of solar and atmospheric neutrino as well as that of neutrinos generated by sterile neutrino DM decay. }
\label{fig:results-nucl}
\end{figure}
Now, we describe the prospect of the signals from the sterile neutrino decay in direct DM detection experiments.
The expected event rate with its spectrum in nucleus recoil scattering off a Xenon nucleus is shown in Fig.~\ref{fig:results-nucl}. 
In this plot, the benchmark points are same as in Fig.~\ref{fig:sk-bound} and
the masses and coupling are taken to be the same values in Figs.~\ref{fig:mixing-mass} and \ref{fig:sk-bound}.
The black and purple curves stand for the contribution of so-called neutrino floor induced by solar and  atmospheric neutrino, respectively.   
We find for $\tau_{\nu_s} \lesssim 10^{21}$ second, the predicted event rate is larger than those by atmospheric neutrinos.

\section{Conclusion} \label{sec:conclusion}

We have investigated the possibility of indirect search of sterile neutrino DM in direct DM search experiments at deep underground.
An extra $U(1)$ interacting sterile neutrino DM, at the tree level, can decay into three lighter neutrinos via on-shell the extra gauge boson decaying into two neutrinos. 
The produced neutrinos are energetic enough to be detected at the direct DM detection experiments.
If $m_{\nu_s}$ is of the order of tens MeV, produced active neutrinos scatter off nucleus. 
If the lifetime is shorter than about $10^{21}$ second, the expected event rate is larger than those by atmospheric neutrinos.
X/gamma-ray searches and other neutrino detection experiments such as SK are complementary to prove this DM scenario.


\section*{Acknowledgments}
We are grateful to J.~Pradler and S.~Matsumoto for enlightening comments.
This work is supported, in part, by JSPS KAKENHI Grant Nos.~JP19K03860 and JP19K03865, and MEXT KAKENHI Grant No.~JP19H05091 (O.~S), 
and JSPS KAKENHI Grant Nos.~JP18K03651,~JP18H01210 and MEXT KAKENHI Grant No.~JP18H05543 (T.~S).

\bibliographystyle{apsrev}
\bibliography{biblio}

\end{document}